# Triangle Mesh Slicing and Contour Construction for Three-Dimensional Printing on a Rotating Mandrel

Kyle Reeser[1], Christopher Conlon[2], Amber L. Doiron[3]

[1]Kyle Reeser, BI2630, Department of Biomedical Engineering, Binghamton University, 4400 Vestal Parkway East, Binghamton, New York, 13902, Tel: +1 585-406-4154
Email: kreeser1@binghamton.edu

[2]Christopher Conlon
Unaffiliated
Tel: +1 585-472-4082
Email: christopher.conlon.mail@gmail.com

[3]Amber L. Doiron, Votey 313, Department of Electrical and Biomedical Engineering, The University of Vermont, 33 Colchester Ave, Burlington, VT, 05405, Tel: +1 607-206-0440
Email: adoiron@binghamton.edu

**Manuscript Keywords:** Additive-Lathe, 3D Printing, 3D Bioprinting, Triangle Mesh Slicing, Cylindrical Printing, Path Planning

**Abstract**

Three-dimensional (3D) printing is a powerful development tool both in industry, as well as in biomedical research. Additive-lathe 3D printing is an emerging sub-class of 3D printing whereby material is layered outward from the surface of a rotating cylindrical mandrel. While established additive manufacturing technologies have developed robust toolpath generation software, additive-lathe publications to date have been relegated to the most basic of proof-of-concept structures. This paper details the theory and implementation of a method for slicing a triangulated surface with a series of concentric, open, right circular cylinders that represents a crucial step in creating toolpaths to print complex models with additive-lathe technology. Valid edge cases are detailed which must be addressed when implementing a cylindrical slicer to produce non-intersecting closed contours; two classes of resultant closed contour are described. Methodologies for generating infill patterns, support structures and other considerations for toolpath construction are required prior to full implementation of a machine capable of printing complex geometry from a digital model onto a rotating cylindrical surface. This work represents the first thorough examination of the mathematics and algorithmic implementation of triangle mesh slicing with concentric cylinders and offers insights for future works in toolpath generation for the additive-lathe type 3D printer.

**Introduction**

*Additive Manufacturing Process Planning*

Three-dimensional (3D) printing is a process by which material is added in a layer-wise fashion to construct a physical object from a digital model. A key consideration in 3D printing is the development of printer-specific software for model processing and toolpath construction.[1] In conventional, rectilinear process planning, a plurality of stacked two-dimensional closed-contours is generated by slicing a 3D stereolithography (STL) model with a set of infinite parallel planes orthogonal to the build direction. An STL file approximates the boundary surface of a digital model as a closed polyhedron via triangle tessellation. Each facet in an STL file appropriate for 3D printing consists of vertices ordered counter-clockwise when viewed from outside of the model, and a unit normal vector pointing outward from the model. Vertices are ordered according to the right-hand rule; this regular ordering scheme allows for the calculation of the unit normal and, in turn, identification of model interior and exterior. Adjacent facets must meet each other along a common edge and share exactly two vertices.[2]

A crucial step while preparing a fused deposition modeling (FDM) device for printing, the user must orient the tessellated model in the virtual workspace of the printer host software. This orientation step serves two main purposes: defining the surface of the model which will be printed onto the flat surface of the build platform, becoming the de facto 'bottom' of the part, and positioning the model with respect to the build direction. The former is important from an aesthetic point-of-view, as the base-layer of a printed structure can be noticeably different in appearance. It is also important from a production standpoint, as a wider, flatter base will help prevent part warp during construction. The latter is crucially important for the success of a print that might have features impossible to print in certain orientations, such as drastic overhanging geometry or voids in the model.

Recent advances in dual-extrusion technology that allow for co-printing of a sacrificial material, and slicing software capable of adding support structures for overhangs has abated many confounding issues with early FDM printers.[3] Still, time and production costs render a host of designs of varying subjective complexity unfeasible.

A 3D printer wherein material is added radially-outward on a rotating mandrel could prove useful in certain cases. The impetus is especially apparent in 3D bioprinting of complex biological structures such as trachea, blood vessels, long bone, and other tissues that include axial voids, often printed with materials which may collapse under their own weight prior to fully crosslinking or require the extensive and time-consuming co-printing of sacrificial material.[4]

*Functional Principle of Additive-Lathe*

Several examples of additive-lathe type 3D printers, wherein the flat build platform of a conventional FDM printer is replaced by a rotating mandrel, the exact angular position of which can be precisely controlled by a stepper motor, have demonstrated the potential of the technology since the first working prototype machine was demonstrated as part of a group engineering project at Imperial College London in 2013.[5] A United States patent has been granted for embodiments of the technology as well, broadly outlining methods that could be used to build and implement such a device.[6] Though wide in breadth, this patent provides limited insight for implementing ground-up model slicing for toolpath generation. Liu and colleagues developed their 'rotary printing device' for bioprinting vessel-like structures,[7] with increasing layers and utility demonstrated by another group.[8] Revotek Co., Ltd. has commercialized their patented 3D blood vessel bioprinting technology, making headlines with the successful implantation of printed vessels into rhesus monkeys.[9] In addition to the potential in vascular tissue engineering, utilizing the technology to print biodegradable stents has been demonstrated.[10-13] The printed structures created with these machines to date, however novel, are limited to only the most basic of geometric complexity.

This paper provides a thorough examination of algorithms and insights which address triangle mesh slicing for 3D printing complex structures on a rotating mandrel. The approaches herein described assume an embodiment of such a 3D printer with a rotational a-axis colinear with the x-axis of the machine. The extruder is assumed free to travel along the gantry in the x-direction, but is fixed in the y-direction. The z-height of the mandrel assembly below the extruder should be positionable for layer-wise addition of printing material (Figure 1a).

*Introduction to Cylindrical Slicing*

Additive manufacturing process planning is a multi-step endeavor; each step is imbued with its own complexities and challenges to overcome, the broadest categories of which are: slicing, contour construction, infill and support structure calculation, and tool path generation.[14] The output of conventional slicing software is a stacked series of closed-loop polygons, ordered according to z-layer height, which is incompatible with rotational printing. A natural analog for a rotational printer would be a concentric series of closed contours, each resting on equidistant imaginary cylinders corresponding to a distance $r$ from the printing mandrel (Figure 1b-e). To enable a rotational slicer plug-in for existing 3D printing toolpath generation software, methods are outlined here to solve the mesh slicing and closed contour construction steps in Cartesian coordinates.

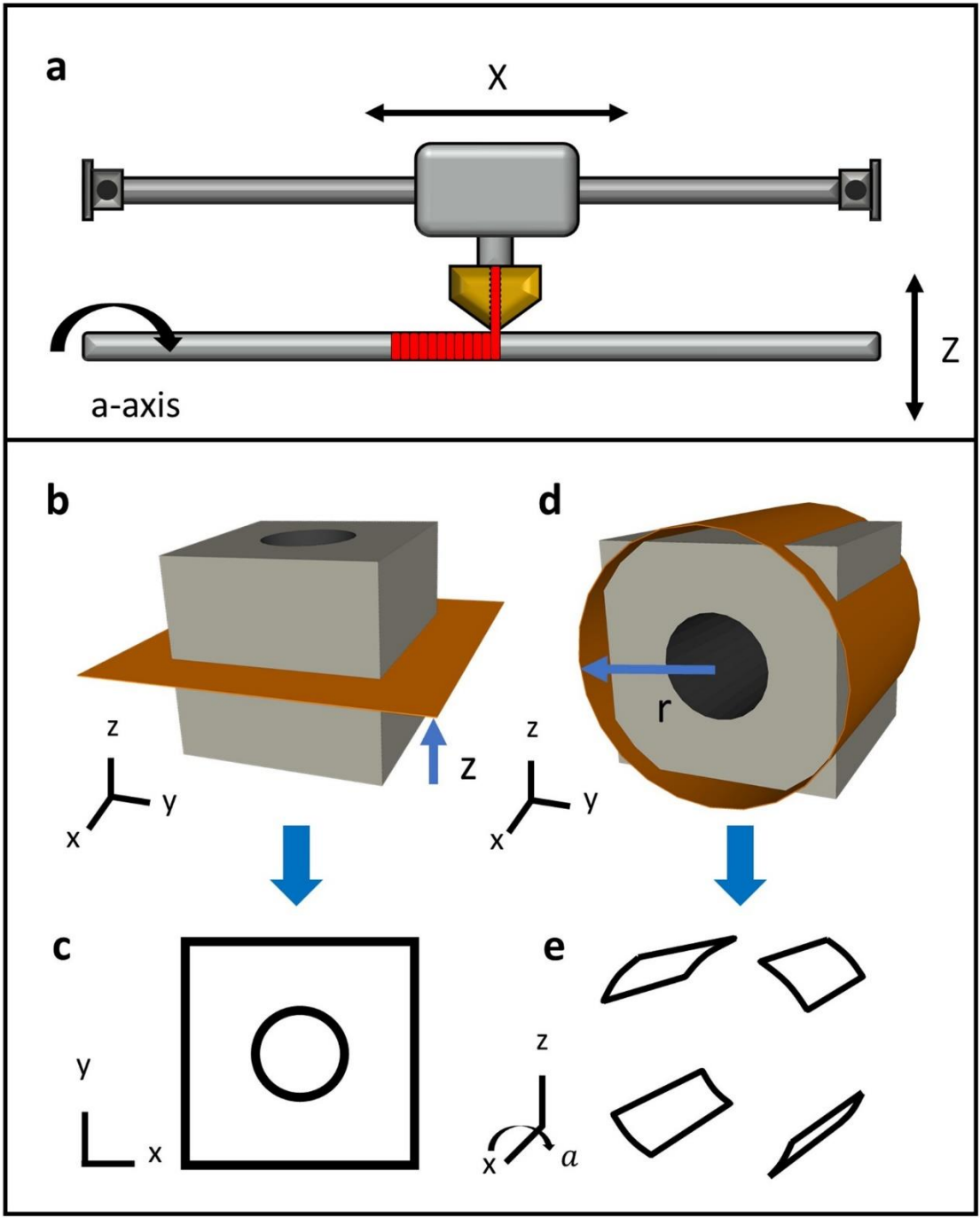

Figure 1: a. Diagram of an embodiment of a cylindrical 3D printer, with movement in the rotating a-direction replacing the y-direction of a Cartesian system. b. A cube with center bore hole prepared for traditional FDM printing. c. A corresponding slice taken from b. d. A cube with center borehole prepared for cylindrical FDM printing. e. A corresponding slicyl taken from d.

**Methods**

*Statement of Problem*

Homologous to the slicing plane in conventional slicers, a slicing cylinder, or slicyl, is herein defined as an infinitely long right cylinder of radius $r_s$ centered along the x-axis, with equation:

$$r_s^2 = y^2 + z^2 \qquad (1)$$

The input of the slicing problem is an unordered list of n facets $F = (F[1], F[2], ..., F[n])$. A set of $k$ slicyls is defined by a list of increasing concentric cylinder radii $r_s = (r_s[1], r_s[2], ..., r_s[k])$ with equal spacing $\delta$ between them.

For $i$ in $k$, slicyl radius $r_s[i]$ of slicyl $s[i]$ is defined as the radius of the mandrel plus an integer multiple of the layer thickness $\delta$,

$$r_s[i] = r_m + i * \delta$$

For each facet $F[j]$, a set of three vertices $F[j]_{V0}$, $F[j]_{V1}$, and $F[j]_{V2}$ is defined, along with a facet normal $F[j]_N$, a unit vector pointing outwards from the model. The output of the intersection problem is a list $A[i]$, containing pairs of intersection point coordinates defining geodesics generated by intersections of all facets with slicyl $s[i]$. A set of $q$ pairwise disjoint closed contours is generated by joining these geodesics such that each item in $A[i]$ is used in exactly one closed contour.

*Model Creation*

A model appropriate for this slicing methodology must be a closed, orientable 2-manifold which divides space into two regions, inside and outside of the model, and should include an axial void which extends completely through the model as a best practice. The mesh is assumed to be watertight and free from any additional topological errors prior to slicing.[15]

*Model Orientation*

The model is imported into the slicer in an arbitrary spatial orientation, from which the triangle mesh is moved en masse to the virtual workspace of the printer. A transformation must be applied to all facet vertices such that the x-axis passes through the axial void of the model, after which a recalculation of unit normal vectors is performed. Moving the triangle mesh prior to slicing not only vastly simplifies subsequent mathematics, but the resulting closed curves generated in the digital workspace map exactly to the physical printer space.

After importing the STL file and generating a triangle mesh, a vector $\vec{v} = \overrightarrow{AB}$ is defined such that the initial point **A** lies at one opening of the axial bore hole, and the terminal point **B** lies at the opposite opening, skewering the model (Figure 2a).

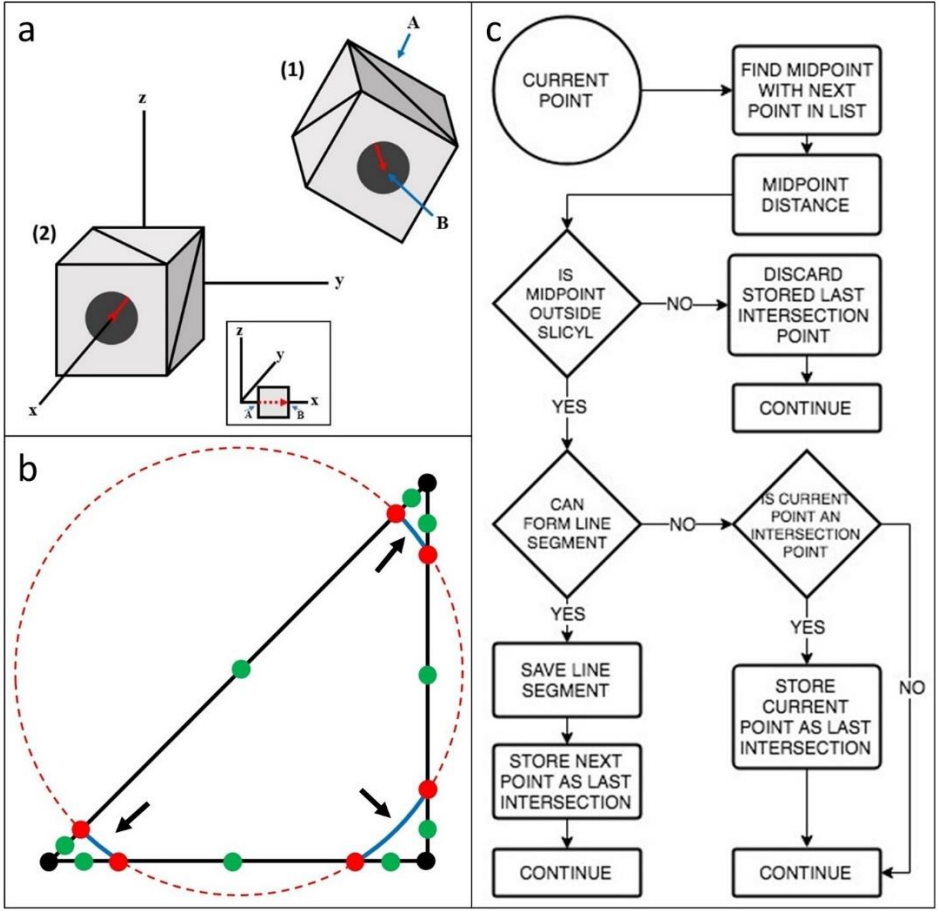

Figure 2: a. A model imported into the slicer in an arbitrary position is transformed to lay along the positive x-axis, with a post-transformation side view (inset). b. Diagram of a slicyl-facet intersection, with associated relevant line segments (black arrows). c. A flowchart describing the process for determining connected intersection points.

The necessary relocation is achieved by applying transformation $M$ to vector $v$, moving all facet vertices relative to vector $v$, with:

$$M = RT$$

where first the translation $T = (x, y, z) \rightarrow (x + \Delta x, y + \Delta y, z + \Delta z)$ is applied to each vertex in the mesh such that **A** is translated to the origin, followed by rotation **R** with **A** anchored to the origin, such that **B** falls on the x-axis ($v_y = v_z = 0$):

$$R = R_y(\theta)R_z(\emptyset) = \begin{bmatrix} \cos\theta & 0 & \sin\theta \\ 0 & 1 & 0 \\ -\sin\theta & 0 & \cos\theta \end{bmatrix} \begin{bmatrix} \cos\emptyset & -\sin\emptyset & 0 \\ \sin\emptyset & \cos\emptyset & 0 \\ 0 & 0 & 1 \end{bmatrix}$$

If the rotation matrices are applied in this order, the mesh will first be rotated with respect to the z-axis until $v$ lies in the XZ plane, where $\emptyset$ is the angle between the projection of $v$ onto the XY plane and the positive x-axis:

$$\phi = -atan2(y, x)$$

The $atan2$ function allows for calculation of the quadrant of the output angle from the signs of the inputs. The mesh will then be rotated about the y-axis until $v$ lies on the positive x-axis, where $\theta$ is the angle between the projection of $v$ onto the XZ plane and the positive x-axis:

$$\theta = \frac{\pi}{2} - \cos^{-1}\left(\frac{p_z}{\sqrt{x^2 + y^2 + z^2}}\right), 0 \leq \theta' \leq \pi$$

All points in the triangle mesh are thusly transformed such that they maintain their spatial orientations with respect to vector $v$, and by extension the centerline of the axial void, coincident with the x-axis. A subsequent step may be taken to translate the model in the positive x-direction for optimal positioning in the virtual workspace of the printer. It is unnecessary to rotate the model about the x-axis prior to slicing.

*Prevention of Slicyl-Vertex Intersection*

The STL file type allows for facets to share a coincident vertex with many adjacent facets.[2] A slicyl-vertex intersection could therefore lead to redundancies that can introduce exceptions into loop-closure algorithms. This work assumes that the values of $r_s$ differ from the distance $d(V)$ of all facet vertices to the x-axis, where:

$$d(V) = \sqrt{V_y^2 + V_z^2}$$

And $V_y$ and $V_z$ are the y and z components of a vertex in the facet $F[j]$.

A slight alteration of facet vertex coordinates or slicyl radii to avoid slicyl-vertex intersection is justifiable due in part to the physical resolution limits of FDM 3D printers.[16]

*Calculation of Bounding Cylinder*

Analogous to the conventional bounding box, a bounding cylinder is herein defined as the smallest, finite, axially-oriented right cylinder centered along the x-axis, which fully encompasses the 3D triangular mesh after transformation. The bounding cylinder is defined by a cylinder radius $r_{BC} = d(V_{max})$, derived from the facet vertex in the mesh farthest from the x-axis, and a cylinder length $l_{BC} = V_{xmax} - V_{xmin}$, derived from the minimum and maximum vertex x-values in the mesh.

*Calculation of Number of Slicyls*

The number of equally-spaced slicyls $k$ that the model will be divided into is contingent on the starting radius of the mandrel, the size and orientation of the model, and the user-defined layer thickness, $\delta$.

$$k = floor\left(\frac{r_{BC} - r_m}{\delta}\right)$$

The first slicyl to intersect the model is set to a radius equal to the mandrel radius plus one layer-thickness, $r_m + \delta$. Alternatively, an adaptive slicing mechanism could be implemented.[17]

*Mesh Slicing: Calculating Intersection Points Between Triangle Mesh and Slicyl*

For each facet in the triangle mesh, points of any intersections of each facet edge with slicyl $s[i]$ are calculated, where each edge is bound by a pair of facet vertices. Parametric equations for a line defined by two vertices in $F[j]$, say $F[j]_{V0} = (x_0, y_0, z_0)$ and $F[j]_{V1} = (x_1, y_1, z_1)$, are:

$$x = x_0 + t(x_1 - x_0) \tag{2}$$
$$y = y_0 + t(y_1 - y_0) \tag{3}$$
$$z = z_0 + t(z_1 - z_0) \tag{4}$$

containing the facet edge in the range $0 \leq t \leq 1$.

To calculate points of intersection, (3) and (4) are substituted into the equation of the slicyl (1):

$$r_s^2 = (y_0 + vt)^2 + (z_0 + wt)^2$$

where $v = y_1 - y_0$, and $w = z_1 - z_0$.

Expanding the terms and rearranging yields a quadratic equation in $t$:

$$At^2 + Bt + C = 0$$

with coefficients: $A = v^2 + w^2$, $B = 2(y_0 v + z_0 w)$, and $C = y_0^2 + z_0^2 - r_s^2$.

Solving for $t$, one or two real roots indicate a corresponding number of intersections between the infinite line and the slicyl, but not necessarily within the line segment that corresponds to the facet edge of interest. Once the roots are calculated, the potential intersections must be sorted for those that fall in the range $0 < t < 1$ on the parametric line. For all facet-slicyl intersections retained, the associated $t$ values are substituted back into (2)-(4) to calculate the Cartesian coordinates of intersection between the current slicyl and the triangular mesh.

*Mesh Slicing: Connecting Relevant Intersection Points*

After determining the points where slicyl $s[i]$ intersects the triangle mesh, it is necessary to determine which pairs of adjacent intersection points connect to each other, and which do not. Only those pairs of intersection points between which the slicyl passed inside of the facet contain information about the model, and are thusly retained (Figure 2b-c).

For each facet, an ordered list of vertices (Figure 2b, black dot) and intersection points (Figure 2b, red dot) along the edges between them is generated, and a midpoint (green) between each intersection and adjacent intersection or vertex is calculated. Beginning with an arbitrary CURRENT POINT in the ordered list (Figure 2c), the distance between the following midpoint and the x-axis is calculated. If the midpoint distance is less than the slicyl radius, the arc of the slicyl passes outside of the facet and contains no relevant information; the next point in the list becomes CURRENT POINT. If (a) the midpoint distance is greater than or equal to the slicyl radius, (b) CURRENT POINT is an intersection point, and (c) there is a stored LAST INTERSECTION POINT, then a relevant line segment has been found, and is saved. If (a) and (b) hold true but there is no stored LAST INTERSECTION POINT, CURRENT POINT is stored as LAST INTERSECTION POINT, and the cycle repeats.

*Mesh Slicing: Contour Construction*

An implementation of a hash-based algorithm proposed by Minetto et al. was adopted for contour construction.[17] In brief, the algorithm utilizes a hash table composed of points that intersect the slicyl. Each key in the hash table is an intersection point, with a corresponding value composed of a list of two adjacent points that form conjoint line segments to either side of the key point. Segments are constructed by choosing a random start key in the hash table and removing its two values after arbitrarily assigning each to either next_point or final_point. At the index of key next_point, one of the two corresponding values is confirmed to be the previous key, the other becomes the subsequent next_point, and both are removed from the hash table. This process continues until no values are found in the table at the index of next_point, at which time the contour is closed after adding final_point assigned from the first key. If the hash table contains more than one closed contour, the cycle repeats. Choi and Kwok proposed a head-to-tail search mechanism for contour construction that could alternatively be implemented.[18]

Closed-curves generated during this step define the boundaries of regions where infill should be placed. We note the existence of two distinct types of closed contour as a consequence of rotational slicing: (type I) a closed curve that lays on the slicyl like a tattoo on an arm, analogous to closed-loop polygons in conventional slicing, and (type II) a closed circumferential curve which winds completely around a slicyl, like a ring on a finger (Figure 3b). Closed circumferential curves will always manifest in disjoint pairs, where infill should be added between them.

## *Implementation, Validation, and Optimization of the Method*

### *List of Active Facets*

A complex STL file may include many thousands of facets, and when combined with slicing at fine resolution can lead to excessive slicing times.[19] Initiating a hash map that stores slicyl indices as keys and a list of facets that will intersect the slicyl as values eliminates the need to iterate over the entire unsorted list of facets for each slicyl, saving computation time.[20] Furthermore, this pre-calculation of a list of active facets screens out potential edge cases, as slicing unsorted facets may yield keys in the hash table that have fewer or greater than two values (Figure 3c-h).

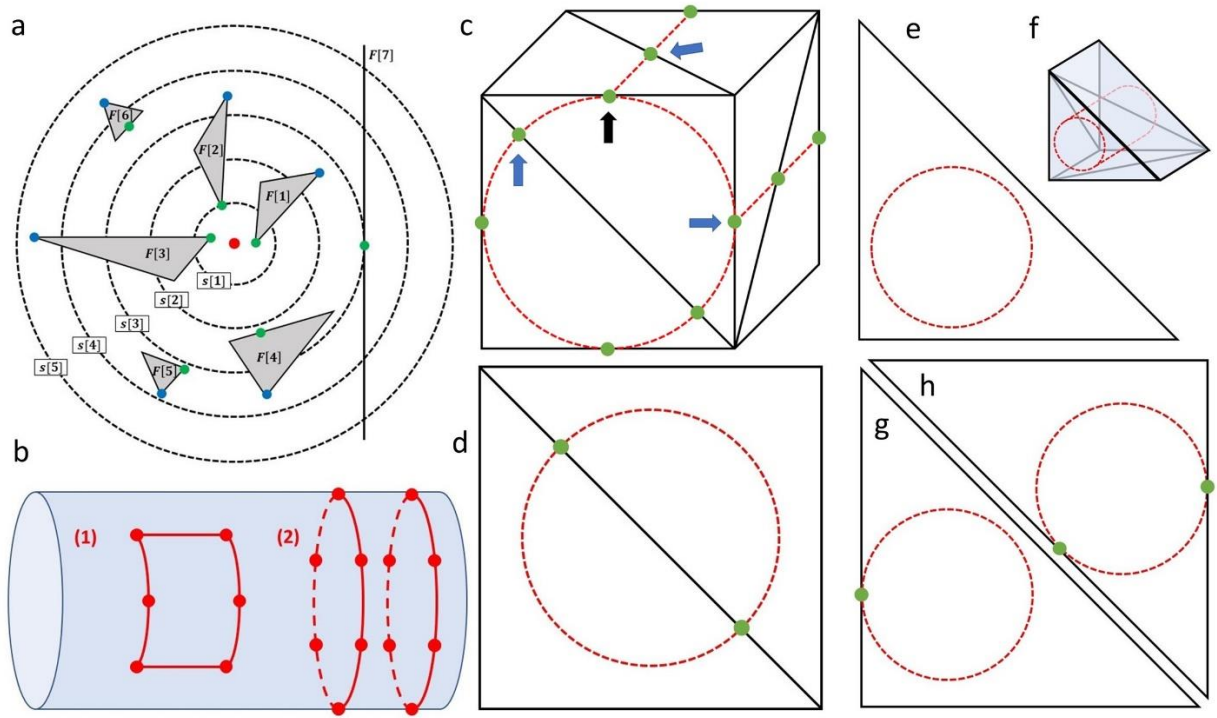

Figure 3: a. An example of active triangle list generation. b. Examples of closed curve (1) and a pair of closed circumferential curves (2). c. An example of an edge case in which an intersection point (black arrow) can be connected to three additional intersection points (blue arrows) which can confound loop-closure algorithms and toolpath generation algorithms d. An example of a valid edge case in which an intersection point is connected to only one other point, generating a closed circumferential curve. e, f. An example in which a slicyl may pass completely through a model with no intersection points. g, h. Examples of edge cases in which a slicyl may intersect a facet at only one or two locations. Edge cases presented in c-g are mitigated through use of the active triangle calculation algorithm and the inclusion of an axial bore hole in the digital model to be sliced.

Facet orientation with respect to the slicyl will dictate $F[j]_{P_{min}}$, the point at which slicyl $s[i]$ might first encounter facet $F[j]$, and this potential tangent point may be at a vertex, or lie along an edge (Figure 3a). $F[j]_{V_{max}}$, the vertex of facet $F[j]$ that is the greatest distance $d$ from the x-axis, will always be the last point in any facet enveloped by the slicyl. Facet $F[j]$ is considered active for slicyl $s[i]$ if and only if $d(F[j]_{P_{min}}) < r_s[i] < d(F[j]_{V_{max}})$.

If slicyl $s[i]$ meets facet $F[j]$ vertex-first, $F[j]_{P_{min}} \equiv F[j]_{V_{min}}$ and is therefore active between $F[j]_{V_{min}}$ and $F[j]_{V_{max}}$. A facet with a normal vector that is not orthogonal to the x-axis and has closest point $F[j]_{P_{min}}$ to the x-axis on an edge will meet a slicyl edge-first. A facet with a normal vector orthogonal to the x-axis may meet slicyl $s[i]$ along a line of tangential contact, generating a curve in the contour construction step, which is closed but non-traversable (Figure 3c).

For a pair of vertices in $F[j]$, say $F[j]_{V0}$ and $F[j]_{V1}$, a line is parameterized according to (2)-(4). The global minimum distance between the parametric line and the x-axis lies at the coordinate $t$ where $r'(t) = 0$ and where

$$r'(t) = \frac{2v(tv + y_0) + 2w(tv + z_0)}{2\sqrt{(tv + y_0)^2 + (tw + z_0)^2}}$$

Setting the derivative to 0 and rearranging,

$$t = \frac{-(vy_0 + wz_0)}{v^2 + w^2}$$

If $t \in (0,1)$, $E_{01_{Pmin}}$, the point on edge $E_{01}$ closest to the x-axis, is given by the evaluation of (2)-(4) at $t$. If $t \notin (0,1)$, the global minimum is outside of the bounds of the facet edge, and $E_{01_{Pmin}}$ is given by $\min\{d(F[j]_{V0}), d(F[j]_{V1})\}$.

This process is repeated over each pair of vertices in facet $F[j]$, and $F[j]_{P_{min}}$ is given by $\min\{E_{01_{Pmin}}, E_{12_{Pmin}}, E_{20_{Pmin}}\}$.

**Results and Discussion**

The slicing and contour construction theory was implemented as a Python package that modularized each step in the process. A dynamic visualizer used for verifying the quality of each slicing step was built as a small web application using node.js to serve content, three.js to render triangle meshes, and raw JavaScript to make the application interactive. As a proof-of-concept, a simple cube was modeled with a hole of radius $r_m$ bored between a pair of opposing faces, corresponding to the radius of the mandrel that would be chosen if physically printing the object. Exported in STL format, the model was sliced with the center of all slicyls coincident with the center of the borehole (Figure 4a). Beginning with a slicyl of radius $r_s[1] = r_m + \delta$, we observe no intersections with the borehole tessellation, as intended (Figure 4b). Until the point at which the diameter of the slicyl exceeds the side length of the cube, all slicyls result in pairs of type II closed contours, where the slicyl intersects with the front and back faces of the model. An infill algorithm should recognize that material is to be added in the space between these two disjoint closed curves. When the slicyl diameter is larger than the side length of the cube model, slicing commences producing four type I closed contours (Figure 4c). With each slicyl, the closed contours become smaller, tapering into the sides and corners of the cube. As in Cartesian FDM, infill material is to be added interior to these closed contours. When physically printed, this model would first manifest as a simple cylinder of increasing thickness with each additional layer applied. The cube would then take form as the four type I closed contours are added with each subsequent layer (Figure 4d).

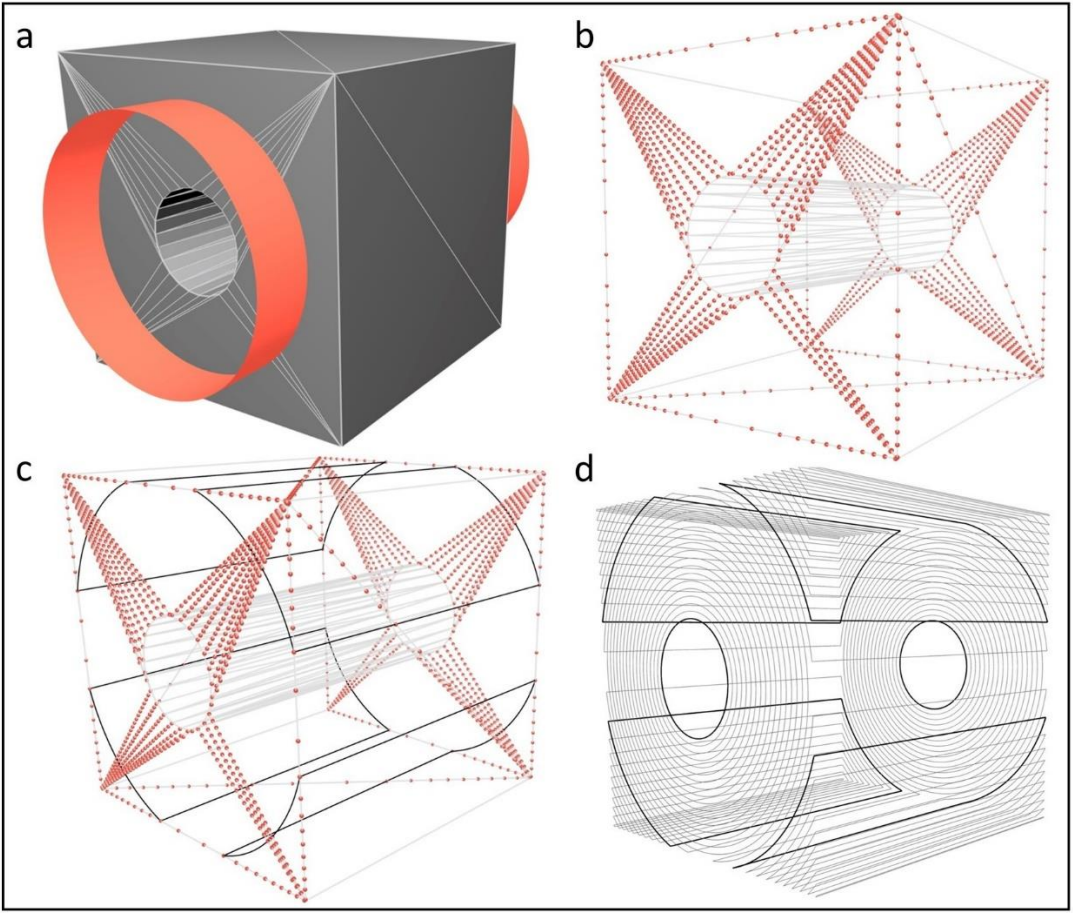

Figure 4: a. Cube model with axial bore hole showing slicyl intersection. b. Tessellated model with overlaid model-slicyl intersections points. c. Tessellated model with overlaid model-slicyl intersection points and one layer of closed contours. d. Fully sliced model with selected layers displaying closed curves (type I), and closed circumferential curves (type II).

A three-blade propeller model[21] was sliced as an additional proof-of-concept, and as a more pointed example of a part one might print with the additive-lathe (Figure 5). Due to severe overhangs and curves, propellers can require extensive support structures with a surface quality affected by stair-stepping. Printing such a part using the additive-lathe could reduce printing time while increasing surface finish quality, making it a prime candidate for this technology.

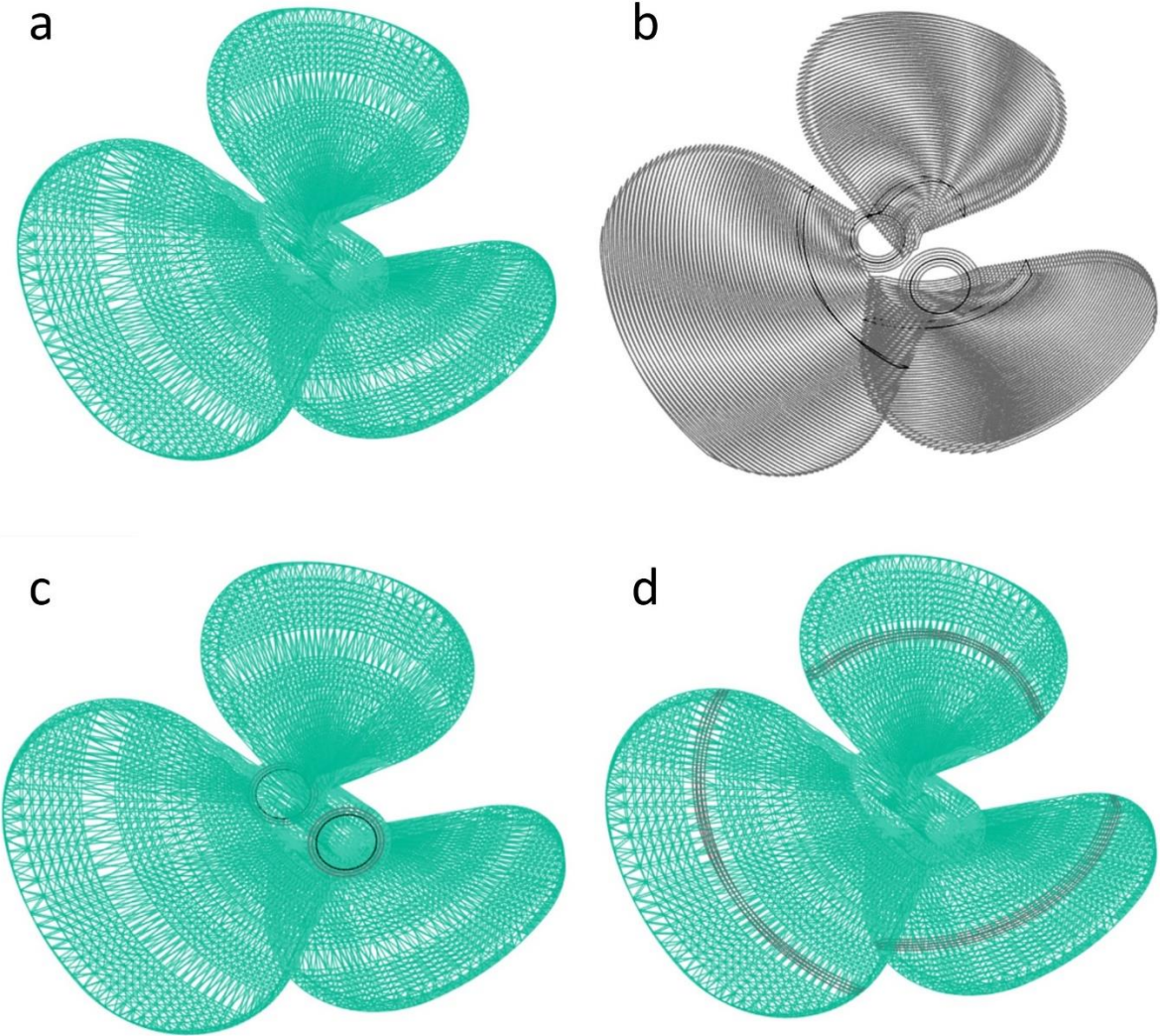

Figure 5: A three-blade propeller model. a. The three-blade propeller model as an STL. b. The sliced three-blade propeller model with layers in the hub and blades highlighted. c. The three-blade propeller model STL with three sliced hub layers overlaid. *Note:* Sliced hub layers are pairs of type II closed contours. d. The three-blade propeller model STL with three sliced blade layers overlaid. *Note:* Sliced blade layers consist of type I closed contours.

Figure 6 shows the fully-sliced model in the top left panel. All constituent layers are shown in subsequent panels, with the current layer paired with the two previous layers for continuity and clarity. Similar to the cube model, the layers begin as a stacked series of type II closed contours to form the propeller hub. When the slicyl diameter is larger than the hub diameter, slicing results in three type I closed contours forming the blades.

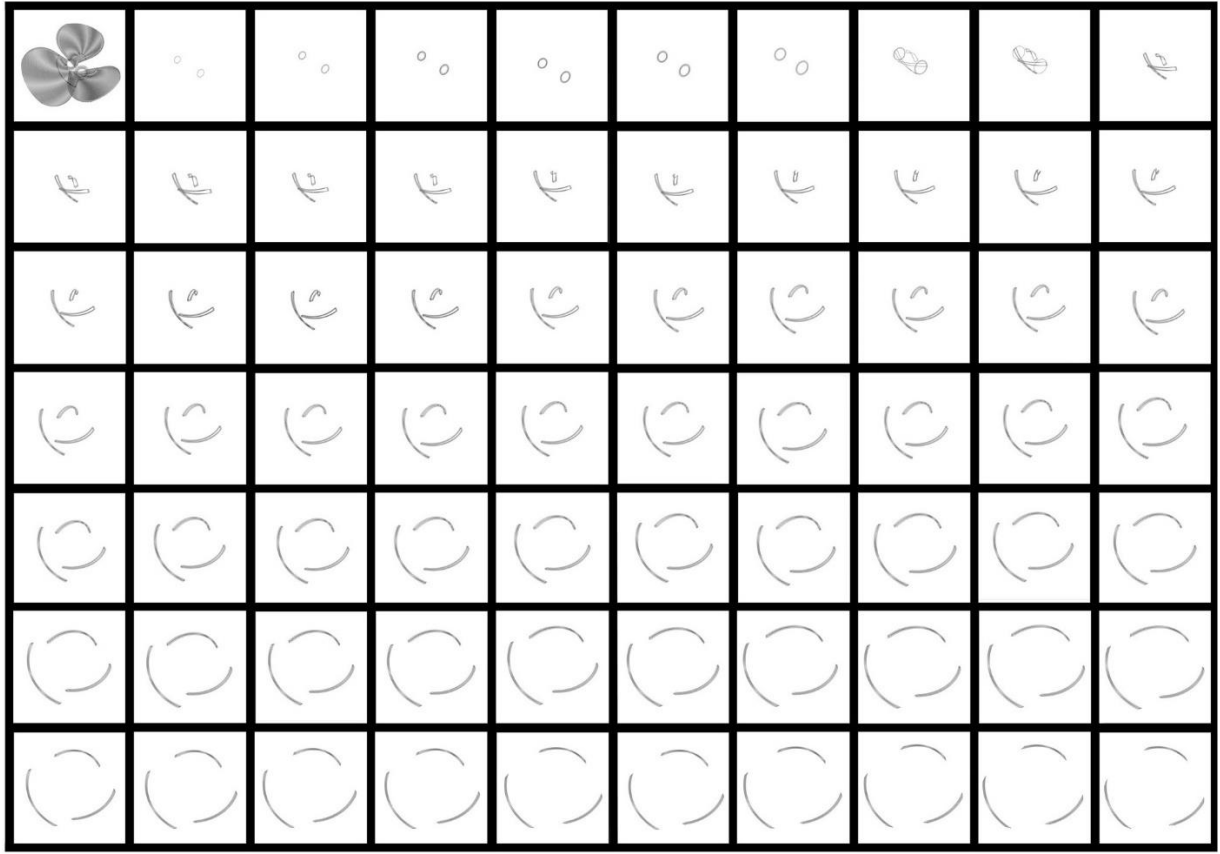

Figure 6: A tiled view of progressive layers resulting from slicing a three-blade propeller model. The fully sliced model is in the upper left tile. Each successive tile represents a cylindrical slice along with the two slices which precede it for continuity.

Cylinders are developable surfaces with zero Gaussian curvature, ergo closed contours calculated in Cartesian coordinates corresponding to each slicyl could be unrolled into a plane without distortion. These flattened contours could potentially be fed into a modified version of existing toolpath generators; with the substitution of the mandrel motor for the y-axis motor on a traditional 3D printer control board and with little additional modification of printer host software and firmware, a "y" coordinate in a movement command could enable a curved extrusion along the mandrel in the a-direction (Figure 1a). Closed circumferential curves will require special care. For a dedicated rotational 3D printer, a native firmware that can process and execute commands in cylindrical coordinates may be developed, but would require a ground-up overhaul of additional process-planning algorithms. For example, algorithms suggested by Volpato and colleagues could be modified and implemented to identify directions of type I closed contours for the purpose of establishing boundaries for infill, but a novel approach is necessary for processing type II closed contours.[22]

It should be noted that models to be sliced with this methodology do not technically require an intrinsic axial borehole; one will be generated during the slicing process based on the location of skewering vector $\boldsymbol{v}$. The tessellation surrounding a pre-designed borehole does however aid in avoiding certain edge cases, such as a slicyl passing entirely within the bounds of a facet as in Figure 3e-f. In such a case, as the slicyl does not intersect with the facet edge, no material would be printed in that location. Passing through a facet without intersection is not a possibility in traditional slicing.

This algorithm assumes the first slicyl to intersect the model is set to a radius equal to the mandrel radius plus one layer-thickness, $r_m + \delta$. Thus, tessellation to the inner wall of a smooth, cylindrical axial borehole with radius $r_m$ will never be active during slicing, with all vertices interior to the slicyl, and can be disregarded. For an axial void with complex inner topography, it will be necessary to design algorithms for calculating the placement of supports or sacrificial material, and thus the axial void tessellation cannot be disregarded. Figure 7 details a model of a lumbar vertebrae, sliced outward from the center of the vertebral foramen.[23] While not radially symmetric, the lumbar vertebrae is a strong candidate for additive-lathe 3D printing as there are no substantial overhangs in the model when printed using this technology.

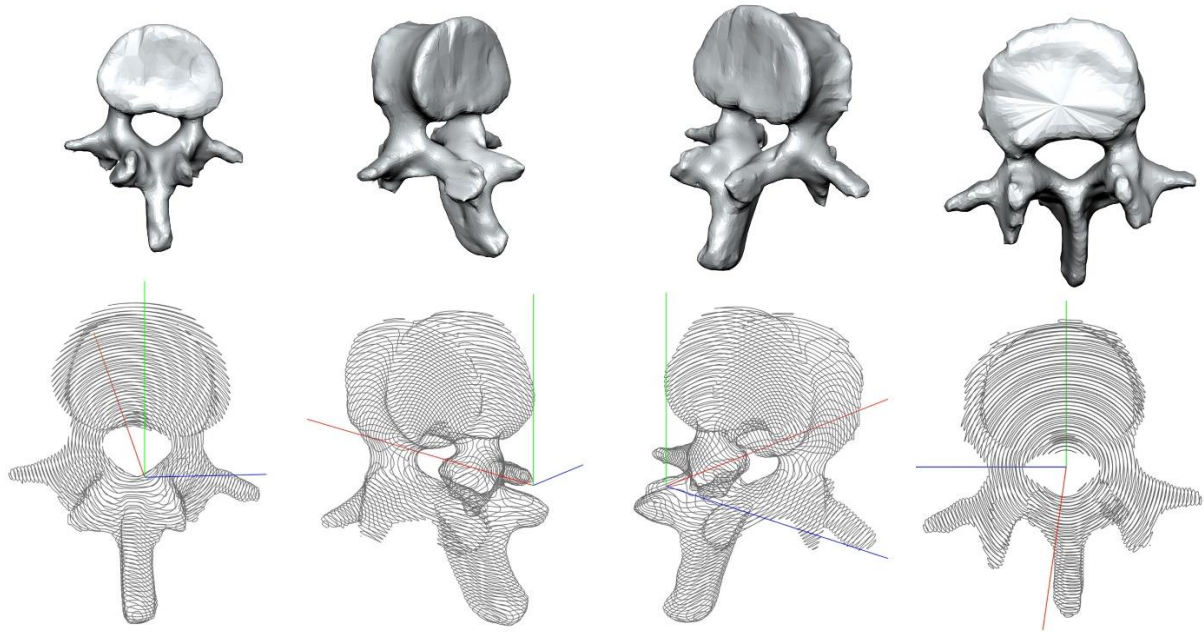

Figure 7: A model of a lumbar vertebrae sliced radially outward from the center of the vertebral foramen. Different orientations of the lumbar vertebrae model (top) are paired with the sliced model in the same orientation (bottom).

**Conclusions**

This paper details slicing and closed contour construction methodologies that are necessary for creating toolpaths to 3D print non-trivial geometry on the surface of a rotating cylindrical mandrel. This work derives methodologies which parallel conventional 3D printing process planning steps, reimagining and expanding in areas for use with an additive-lathe 3D printer. Using slicing cylinders results in a concentric series of closed contours. Generating a list of active triangles for each slicyl allows the slicer to perform more efficiently, while also mitigating potential edge cases that might otherwise confound closed-loop contour construction.

This paper details a slicing methodology that may be utilized in an additive-lathe printer plug-in for conventional, rectilinear 3D printing systems, and if integrated properly allows for incorporating myriad advancements already made in the field of toolpath generation. While the slicing and contour construction steps outlined in this paper are a crucial first step, additional algorithms for infill and toolpath generation are required to realize the full potential of additive-lathe three-dimensional printing technology.

**Acknowledgments:** The authors report no outside acknowledgements.

**Author Disclosure Statement:** No competing financial interests exist.

**Corresponding Author:**

Amber L. Doiron, Votey 313, Department of Electrical and Biomedical Engineering, The University of Vermont, 33 Colchester Ave, Burlington, VT, 05405, Tel: +1 607-206-0440
Email: adoiron@binghamton.edu